\documentclass[12pt]{article}

\parskip      \medskipamount
\usepackage{times}

\begin{document}

\begin{center}
{\bf \large The SETI Episode in the 1967 Discovery of Pulsars}

\vspace*{3mm}
Alan Penny, School of Physics and Astronomy, University of St Andrews

\vspace*{1mm}

Published in European Physical Journal H, February 2013 \hspace*{5mm} www.epj.org \newline
http://dx.doi.org/10.1140/epjh/e2012-30052-6 

\end{center}

\section{ ABSTRACT}

In the winter of 1967 Cambridge radio astronomers discovered a new type of radio source of such an artificial seeming nature that for a few weeks some members of the group had to seriously consider whether they had discovered an extraterrestrial intelligence. Although their investigations lead them to a natural explanation (they had discovered pulsars), they had discussed the implications if it was indeed an artificial source: how to verify such a conclusion and how to announce it, and whether such a discovery might be dangerous. In this they presaged many of the components of the SETI Detection Protocols and the proposed Reply Protocols  which have been used to guide the responses of groups dealing with the detection of an extraterrestrial intelligence. These Protocols were only established some twenty five years later in the 1990s and 2000s. Using contemporary and near-contemporary documentation and later recollections, this paper discusses in detail what happened that winter.

\section{INTRODUCTION}

In the winter of 1967 radio astronomers in the University of Cambridge became perhaps the first group actually to face the `Contact' problems which the SETI (Search for Extraterrestrial Intelligence) community is still grappling with in its Detection and Reply Protocols. If you think you have discovered evidence for the existence of extraterrestrial intelligence, how do you confirm it, and then how do you announce it? Is it then dangerous to send out a signal to such an intelligence?

The group had discovered an entirely new kind of radio source and in two months of detailed investigations determined that they had found the long predicted neutron stars - a discovery that was to be awarded a Nobel Prize. Such was the novelty of the source that its nature was at first completely uncertain and for about three weeks in those two months some of the group leaders had to deal with the possibility that they had detected signals from an alien civilization. Although this never became the preferred explanation, it was seriously considered and one stage the group leaders discussed what their strategies would be in the event they had to announce to the world such a discovery and even whether it might rather be better to keep it secret.

\section{THE DISCOVERY OF PULSARS}

In July 1967 a new low-frequency radio telescope started working at the Lord's Bridge station of the Mullard Radio Astronomy Observatory (MRAO) of the University of Cambridge. Antony Hewish  had led the design and construction of  this novel telescope, a  collection of wooden poles with wires strung between them, built to discover more of the newly found quasars and measure their sizes, by watching them flicker as the Interplanetary Medium passed in front of them. Covering two hectares, this was the largest telescope then working at this long (4-m) wavelength. All the wires were connected to a central laboratory in such a way so that the telescope had  `beams'  which pointed due south and at a fixed declination in the sky. As the Earth rotated, a circle of sky of that declination swept through each beam each day. There were at first three beams covering three declinations. The declinations could be changed by changing the cabling in the system. To see the fast flickering (`scintillation') of a source during the four minutes it spent passing through  the beam, the system had been designed with short integration time recorders, unusual then in radio telescopes. 

Jocelyn Bell (now Professor Dame Jocelyn Bell Burnell, FRS), Hewish's graduate student, had participated in the two years of construction of the telescope and from July was operating the telescope and doing the analysis of strip chart recordings. In addition to detecting quasars and other astronomical sources and many types of radio interference from Earth sources, soon after the start  Bell noticed a source which gave an unusual flickering pattern, which she described as `scruff'. After a few weeks she realised that this source, which looked neither like the other astronomical sources nor like the terrestrial sources of radio interference, sometimes reappeared when the telescope looked in a particular direction. Careful work showed that it was seen at the same sidereal time each day (although not every day), the characteristic of a fixed astronomical source. Sometime in late August she spoke to Hewish about this and it was decided to look at the flickering more closely. By the start of November a second faster chart recorder was then installed. After a month of no-shows, on about November 28th the source reappeared and was revealed as a series of short (less than 0.3 sec) pulses separated by about 1.3 seconds. This was something entirely new in astronomy. The large size, fast response and wavelength of the telescope had opened up a new field of astronomy. No known type of object could behave like this. Checks were made for man-made interference, even including enquiring on November 29th  (Hewish 1967) of nearby astronomical observatories whether they were making radio emissions at a fixed sidereal time. The pulse separation was extremely stable, and since the shortness of the pulses meant that the source had to be smaller than a star, this was clearly a new type of astronomical source. After further work including the use of another telescope and the discovery of three more sources, a plausible explanation for the source could be made and a paper (Hewish et al. 1968) with Hewish and Bell as first and second authors was sent to the science journal {\it Nature} on about February 8th. The discovery was announced in a seminar on February 20th, preceding the appearance of the paper in  {\it Nature}  on February 24th 1968. These announcements included the suggestion that the sources could be vibrating neutron stars. These `pulsars', as they were soon dubbed, were later explained as rotating neutron stars. Hewish was to share in a Nobel prize for his ``decisive role in the discovery of pulsars" (R Swe Acad Sci 2012). The exact sequence of events in the discovery process as recollected by the participants has some divergences and several strands of the discovery and of the subsequent investigation seem to have overlapped (Woolgar 1976).

\section{THE LITTLE GREEN MEN}

The alien civilization aspect of this dramatic discovery developed from the start. Bell marked up the daily 98-feet long strip charts in the attic shared by the graduate students (Mackay, 2008). After she first noticed the `scruff' sometime in August, she then noticed probably in late August that it repeated at the same sidereal time. The source was flickering unusually fast - much like an artificial signal - and was clearly distant. It was almost immediately in September (Wade 1975) that some of the group (who had watched Bell working with the charts) dubbed it `the LGM star', the `Little Green Men' nickname that was to last for some time. Bell wrote in her thesis (Bell 1968)  ``The possibility that the signals were from some intelligent civilisation in the universe was not ruled out: hence the unfortunate nickname `little green men' ". Bell Burnell later commented (Bell Burnell 2011) ``It is always at the back of radio astronomers' minds that they might stumble across signals from other civilizations". As John Pilkington, third author on the discovery paper, now comments (Pilkington 2011) ``We were using the resources that we had to examine a weak and unreliable signal that could have been interference, or equipment malfunction, or a previously unrecognised `natural' phenomenon, or LGM [Little Green Men]. They [LGM] were always a possibility, with awesome implications, but they didn't really drive the investigation." In her 1975 ``Petit Four" speech (Bell Burnell 1977) Bell Burnell said ``We did not really believe that we had picked up signals from an alien civilization, but obviously the idea crossed our minds, and we had no proof that it was an entirely natural radio emission."  However,  Paul Scott (fourth author on the discovery paper) now comments (Scott 2011) ``It [LGM] really just represented an umbrella term for any non-natural source of the radiation.", much as the term `gremlins' is sometimes used to describe the cause of malfunctioning equipment. However when Hewish wrote (Hewish 1967) on 29th November to Donald Lynden-Bell  at the Royal Greenwich Observatory, saying ``I am puzzled by signals from little green men", although this was meant jokingly (Hewish 2012), it does seem to use the nickname to refer to the extraterrestrial intelligence aspect.

But after the observation on November 28th, when it was seen that the shortness of the pulses and their regular timing were clearly unlike any other known source, the explanation that they were from an alien civilization started to become more real. Martin Ryle, the MRAO group leader, later wrote (Ryle 1968b)  that ``Our first idea was that other intelligent beings were trying to establish contact with us." As the investigation continued more `alien civilization'-like aspects emerged.  The pulses varied in strength and were sometimes absent for long periods of time, unusual for variations in an astronomical source. Then an investigation early in December by Pilkington found  that the signal only occurred in a narrow frequency band, which swept downwards in frequency - a nature not then seen in natural sources and much like the behaviour of military radars.  Additionally, the narrowness of the band and the sweep rate meant the source had to be smaller than about 5000 km (Hewish et al. 1968). By mid-December the alien explanation was moving up. Hewish later wrote (Hewish 1968b)  ``As the days went by excitement rose when we found that the pulses were coming a body no larger than a planet situated relatively  close to us among the nearest stars of the galaxy. Were the pulses some kind of message from an alien civilisation? This possibility was also entertained for lack of an obvious natural explanation for signals that seemed so artificial." and recounts in his Cambridge autobiographical recordings (Hewish 2008) that ``all kinds of thoughts went through our minds; it was such an artificial signal that I had to seriously consider that the signal was being sent to us ... I had a test for the little green men idea though I had to pinch myself to take it seriously ...  I was having lunch there [Churchill College]  one day sitting next to Sir Edward Bullard, the geophysicist, and told him; he said that if they were narrow band they were probably intelligence; as he took it seriously I had to too; we were already measuring the band width and they were indeed narrow band but my argument was that if you have intelligent signals they are likely to be coming from a planet, and that would have to be in orbit around a star". Hewish started these timing measures on about December 11th.

Pilkington comments (Pilkington 2011)  that Ryle considered LGMs a ``real possibility". Ryle himself said (Ryle 1971) later in 1971 ``I don't think any of us really believed it was explanation - but it was something we had to think about." In his 1975 Nobel lecture, Hewish says (Hewish 1992) ``We had to face the possibility that the signals were, indeed, generated on a planet circling some distant star,and that they were artificial. I knew that timing measurements, if continued for a few weeks, would reveal any orbital motion of the source as a Doppler shift, and I felt compelled to maintain a curtain of silence until this result was known with some certainty. Without doubt, those weeks in December 1967 were the most exciting in my life.".  It is unclear to what extent the whole group shared in these matters. Hewish states (Hewish 2012) ``We had no group discussions about the LGM possibility", and Scott also notes (Scott 2011) that ``I was a relatively junior member of staff, or course, and would not have been involved in any `high-level' discussions". Craig Mackay, who had a desk next to Bell during the discovery, now comments (Mackay 2012) that the alien intelligence interpretation was ``[talked about] seriously in the earlier stages", although adding ``I do feel that the LGM part of the whole discovery has become rather overblown".

During this time there were a number of aspects to consider if the ongoing investigations showed the alien explanation was to be preferred. First, it was clear that news, even of the fact that an LGM explanation was being considered, should not leak out - a concern expressed by Ryle in a letter (Ryle 1968a) on 9th February to the Astronomer Royal, Sir Richard Woolley, at the Royal Greenwich Observatory. Apart from the sensational media coverage that was bound to ensue, the work of investigation would be hampered, as had occurred before (Hewish 2008). However the discovery process had in any case already been kept confined within the involved members of the MRAO. This is a normal procedure for scientists with a hot discovery, where one wants to be certain before making an announcement and one does not want other people getting wind of the news and leaping in and distracting attention from what was the result of years of effort. But the MRAO group in particular had had in the past problems of this sort with both observational and theoretical colleagues. In particular relations were poor (Mitton 2011) with Fred Hoyle (later Sir Fred Hoyle), a theoretician in another Cambridge department. The news was kept so tight that Longair recounts (Longair  2011) how, although his office was next door to Hewish and that he was a member of MRAO, he knew nothing about even the fact of a source discovery until the announcement seminar on February 20th. 

Then there was the matter of how should any announcement be made. Hewish says (Hewish 2008) ``I had discussed with Martin Ryle before Christmas when I did not know the answer what on earth we were going to do with this data if it turned out that intelligent signals were a likely explanation; you can't just publish it or release it like a news flash; we thought we would inform the Royal Society and get it handled nationally as it was too big a thing to deal with ourselves". Bell Burnell wrote about this meeting in her 1975 speech (Bell Burnell 1977) ``Just before Christmas I went to see Tony Hewish about something and walked into a high-level conference about how to present these results. We did not really believe that we had picked up signals from another civilization, but obviously the idea had crossed our minds and we had no proof that it was an entirely natural radio emission. It is an interesting problem Ñ if one thinks one may have detected life elsewhere in the universe how does one announce the results responsibly? Who does one tell first? We did not solve the problem that afternoon, and I went home that evening very cross Ñ here was I trying to get a Ph.D. out of a new technique, and some silly lot of little green men had to choose my aerial and my frequency to communicate with us.".  (Bell was in the final year of her PhD.) Bell Burnell later noted (Bell Burnell and Hewish, 2010) that there was `one other senior person' with Hewish and Ryle, although Hewish has no memory of this (Hewish 2012). Possible names for this possible person include Peter Scheuer and John Shakeshaft. Mackay comments (Mackay 2012) that in 1967 he thought that if LGMs became the preferred interpretation then ``the involvement of the Royal Society and very likely [the] government itself in any announcement would be essential".

Finally, should any announcement be made at all? Hewish recounts (Hewish 2008) that at that pre-Christmas discussion with Ryle,  ``[Ryle] was half joking but said burn the records and forget about this, because É if the news gets out that there is intelligence out there É people will want to launch a signal in that direction to talk to them. ... supposing ... [the aliens] are looking for a planet to occupy É the next thing that will happen is that we will be invaded".  Pilkington comments (Pilkington 2011) that ``[Ryle] was concerned that attempts to contact the LGM might lead to disaster" but  adds that  ``[this] was more likely to be an emotional outburst than a reasoned proposal". Certainly this was not discussed with the wider group. Judy Bailey, then a research student, is recorded (Woolgar 1979) by Woolgar in 1975 as saying ``there was  a meeting of the central group and somebody said `Jocelyn's discovered some little green men but you mustn't tell anybody' ... people concerned with that instrument had a meeting and were sworn to secrecy and one of them wasn't very good at keeping his oath so I heard about it." Woolgar also reports that Bell Burnell told him that ``I got quite specifically told on that occasion to act dumb over Christmas." Ryle himself was to say (Ryle 1971) in 1971, when he was talking about the discovery process ``The evidence of our own planet suggests that if you live, for example, in SE Asia the less contact you have with higher civilisations the better. So perhaps we should burn our own records and forget all about it.". In a letter (Ryle 1977) to an enquiring member of the public in 1977 he wrote ``I do think it wrong that one group of scientists in one country should plan to send out very powerful radio signals in an attempt to make such contact, without there having been a proper international discussion.".  Ryle was known (Graham-Smith 1984) to feel strongly about the effects of science on society.

Such a signal was in fact sent out by Frank Drake in 1974 (staff NAIC 1975, Cornell News 1999) and Ryle wrote  (Tarter 2009) to Drake complaining that it was  ``very hazardous to reveal our existence and location to the Galaxy; for all we know, any creatures out there might be malevolent - or hungry". Later, it seems (Sullivan 1976, Lovell 1977) that Ryle led an approach by several people to Sir Bernard Lovell of Jodrell Bank fame who then sent a private letter (Lovell 1976) to the International Astronomical Union raising the possibility of malevolent aliens, saying that  ``I have been asked to seek a discussion in the Executive Committee ... astronomers are involved in the problem of communication with extraterrestrial communities. Transmissions for this purpose are being made .... [ as to whether] the IAU should draw the attention of world governments to a problem which could conceivably be of critical importance'' and ``whether the astronomical community should take steps to initiate a wider discussion on an international basis of the consequences of  success ... I repeat I raise this issue on behalf of a number of distinguished individuals''. After consulting Drake, the IAU concluded (Muller 1976) that no action was needed. It has been said  (see e.g. Norris 2004, Lemarchand and Tarter 1994) that Ryle himself wrote to the IAU, but there is no such letter in the Ryle or IAU archives.

It has been an as yet unresolved matter of much discussion (see e.g. Penny 2012, Haqq-Misra et al. 2012) about whether the dangers of signalling are real. But it appears that Ryle may have been more serious than Hewish realised in raising the possibility of secrecy.

{\small
\begin{table}[!h]
\begin{tabular}{p{1.3in}p{4in}}
Time & Event \\ \hline
     July         &   telescope starts working \\
     Aug 6      &   Bell notes `scruff' \\
     Aug late  &   Bell sees that the `scruff' repeats at the same sidereal time \\
     Aug late  &  Bell speaks to Hewish   \\
     Sep         &   `LGM' nickname adopted jokingly \\
    Nov $\sim$1  &  fast recording started -- source disappears \\
    Nov $\sim$ 28    &  source reappears,  pulses recognised \\
    Nov 29    &  detailed investigations starts \\
    Nov 29    &  Hewish writes to Lynden-Bell `plagued by LGM' (jokingly)  \\
   Dec           & 4C array aerial confirms source \\
   Dec  $\sim$1-12   &   Pilkington finds narrow band sweeping \\
   Dec  $\sim$1-7      &     Hewish talks to Bullard -  starts to think seriously about ET \\
   Dec 11- Jan 5    &    Hewish looks at timing for orbital motion \\
   Dec  21   &   Hewish and Ryle discuss options if it is ET \\
   Dec  21    &   Bell finds second source \\
   Dec          &   Scheuer explains band sweeping/disappearances as ISM \\
   Jan $\sim$1-7  &  ET possibility dropped \\
   Jan   early   &  paper writing starts \\
   Jan  mid   & Hewish conceives neutron star explanation \\
   Jan  mid  & Bell finds third and fourth sources \\
   Feb  $\sim$8 &  paper sent off \\
   Feb 9       &  paper received by Nature \\
   Feb 9       &   Ryle writes to Woolley (mentions ET explanation) \\
   Feb 20     &   announcement seminar \\
   Feb 24     &   paper appears \\
   Feb 24 - March & Hewish, Ryle talk to press about discarded ETs \\
\end{tabular}
\caption{Timeline of events in 1967-1968. Datings are from the references in this paper. There was overlapping of some phases and there are some uncertainties in the exact datings, and so the ordering is to that extent uncertain.}
\end{table}
}

However, these matters were soon to become moot, as the investigations started to point to a natural explanation. Peter Scheuer pointed out (Scheuer 1968) that scintillation in the Interstellar Medium and in any medium around the source would produce the variations in the pulse intensities and the long periods of no-shows. Also the dispersion in the Interstellar Medium, whereby the speed of a radio wave decreases with decreasing frequency, would spread out an initially wide-band pulse, as natural sources at this frequency should be, into the narrow-band frequency sweeping signal observed. Careful timing of the pulses showed no sign of the drift that would be expected if they were from a source on a planet orbiting  star, a likely location for an alien civilization. And more such sources were found - the source was not a singular oddity. Hewish wrote (Hewish 1968b)  ``It [LGM] soon declined in attractiveness with the discovery of similar pulses coming from three other directions in space, and with the absence of any planetary motion associated with the sources. (Presumably another civilization would have to occupy a planet.) .... By this time we felt reasonably confident that the pulsars were a natural phenomenon.", a rationale also given by Ryle in his Feb 9th letter (Ryle 1968a) to Woolley. As Bell Burnell later noted (Bell Burnell 1977) when she discovered the second source in the evening after the Hewish/Ryle discussion, she ``went off, much happier, for Christmas. It was very unlikely that two lots of little green men would both choose the same improbable frequency, and the same time, to try signalling to the same planet Earth." With the lack of orbital motion, multiple sources,  and the variations and frequency sweeping explained,  the pointers towards an artificial source were no longer strong.  Hewish was reported (Woolgar 1979) as saying ``the notion of an intelligent origin had been discounted by the first week of January." The paper writing started at this time. Finally, Hewish conceived (Hewish 2012) of a natural explanation - a neutron star. Neutron stars are the collapsed cores of dead stars and are only ten or so kilometres across. These could vibrate rapidly enough, and the high density of the star would promote stability in the process, although the emission mechanism itself was still to be explained. However, even as late as February 9th when Ryle sent that letter to Woolley, Ryle said there were two possible plausible explanations for the signal, with the first being ``(a)  To suppose that the signals arise from intelligent beings in some other planetary systems", before later saying ``The former possibility can perhaps be discounted". Clearly the LGM possibility was still important enough to be mentioned.

In the published paper no mention was made of the LGM possibility, although this was hinted at by saying that a source with a planetary orbit ``comparable with that of the Earth" was ruled out. But when the reporters arrived, no secret was made (`LGM' had been marked on many of the strip charts) and the nickname became widely known. In the subsequent media coverage, in addition to the relatively more prosaic neutron star explanation, much attention was given to the `LGM' possibility: ``[Ryle] ... it is not impossible - only improbable that the signals came from an intelligent source outside our Solar System" (Reuters); ``The Girl who Spotted the Little Green Men"; ``At first, regular pulsations could only be explained as intelligent signalling" (Daily Telegraph); ``[Ryle] ... at the beginning the breathtaking regularity of the pulses raised the possibility that the signals came from some sort of extra-terrestrial life" (The Times 08 Mar 1968). Many more quotes from the team appeared in other British and overseas papers. Thus the front page of the Daily Mail of 24th February said ``Dr. Anthony Hewish, one of the investigating team, said last night: `At first we thought that there might be intelligence in outer space trying to contact us. We still cannot rule out that possibility'."  It is of interest that at the April 1968 meeting of the Royal Astronomical Society, Hewish is recorded (Hewish 1968a) as saying ``We call the sources CP1919, ... - a more satisfactory nomenclature than L.G.M."

 \section{DISCUSSION}

The concept of a radio signal from an alien civilisation was indeed not a new one at the time. The idea of looking for signals had been raised by Cocconi and Morrison in a 1959 letter to {\it Nature} (Cocconi and Morrison 1959), and an actual search (Drake 1961, 1965) in 1960 at the Green Bank radio observatory had attracted attention (Lawrence 1959) and there had since that time been a number of searches and publications in the scientific literature. In 1960 NASA had published the ``Brookings Report" (Michael 1960a,b) which included a discussion of the effect of discovery of intelligent life, and even an allusion to the possibility of keeping such a discovery secret. The notion of aliens being dangerous had been the subject of a 1961 BBC TV science fiction series (Hoyle and Elliot 1962). Although members of the MRAO group were aware of these things they were generally (Scott 2011,  Hewish 2011) of the opinion that the subject was of little interest. However Ryle had been involved (Ryle 1968b) in discussions about meetings on the search for extraterrestrial intelligence.

Throughout Bell was certainly not enamoured of the LGM idea, seeing it in part (Bell Burnell 1977) as a hinderance in her completing her PhD.  Malcolm Longair, a member of the MRAO group but not involved in this work, recounts (Longair 1996) how Bell sent Hewish a Christmas card which ``consisted entirely of a series of pulses, which, when decoded, contained a Christmas greeting". This was presumably in jest. Bell Burnell now says (Bell Burnell, 2011) ``But the negative approach shows that we were not inclined to suspect this was the correct explanation. However we had to examine it and ideally we had to rule it out before publication." At the present time Bell Burnell strongly disputes (Bell Burnell 2011) the analysis given in this paper. She says that at the time she was not aware of the mid-December conversation between Hewish and Bullard about taking the LGM idea seriously and wonders if this is not a  subsequent gloss. She says that the conversation between Hewish and Ryle was not as reported here, but was merely on how to publish the result given the limited information available at that time. Specifically she has no memory of Ryle discussing keeping the whole thing secret, and thinks that this aspect has only started to be mentioned recently. In fact, the LGM explanation was ``always tongue in cheek, or a joke". ``There never was such a [LGM] `explanation' to be downgraded." Her comment on the analysis given here is that ``spin is the domain of journalists and as scientists we have to respect the data, even if it makes a less exciting story". (To some extent these comments seem to be at variance with her 1968 thesis and her 1975 speech.) Bell Burnell's version is supported to some extent by Scott who comments (Scott 2011) ``I do not think that I - or anyone else - doing this [using the LGM name] thought they were referencing anything other than an astronomical object" and ``I never heard any suggestion of this [keeping it secret]". However the comments recounted here made by Hewish and Ryle at the time and in the years after seem conclusive in indicating that the LGM idea was seriously considered by them at the time.  The Rashomon-like differing accounts seem to derive from the fact that within the group there was not a complete exchange of thoughts at all levels.

Could in fact it been kept secret if an `LGM' had been found? In such a tight-knit (Woolgar 1979, Burbidge 2007, Longair 2011) group the answer is surely yes - we can compare this to the Enigma code breaking of World War II, where a large number of people kept a major matter secret for three decades. In this discovery process no news had leaked out about the discovery before publication, in conformity with the long-established rule in the group that "nothing was said about new results ... until a paper was accepted" (Graham-Smith 1984). In due course another astronomy group would have stumbled on these objects, but perhaps that would have given enough time to set up an international law to ban sending signals out, before that later discovery. There is no record of whether or not Ryle considered this at the time, but his efforts in 1976 to get the IAU to approach national governments shows he had kept the matter in mind enough to take action then.
 
In retrospect, at that time the reasons for downgrading the intelligent signals explanation are in fact not totally conclusive. The lack of orbital motion could be explained  by the aliens putting their source far from a star. The fact that there were a number of these sources would merely mean that aliens were common. The fact that the signals  were intrinsically powerful would just mean that the aliens were very advanced. But overall the judgement was reasonable: there was a plausible natural explanation in pulsating neutron stars and the LGM explanation came solely from the sharpness and regularity of the pulses. It would seem reasonable that if these were actually artificial then there would be more evidence - perhaps there would artificial intensity changes under the intensity variations from the ISM. Any `LGM' claim needs extraordinary evidence, and by early January 1968 that was just not present.

Overall, the process was remarkably fast. From November 28th, when the pulsed nature was first seen to about mid January when the neutron star explanation was adopted during the paper writing was only seven weeks. The `could actually be LGM' period seems to have lasted only three or so weeks from the time of the Bullard conversation until `early January' when the LGMs were dropped. The whole process was a tour de force of persistence, discovery, investigation and theorising.

\section{THE SETI DETECTION PROTOCOLS}

This episode is an early `real-life' example of dealing with concerns of subsequent SETI work: how to test and confirm a SETI discovery, how to announce it, and whether sending a message out is dangerous. The process they went through remarkably presages many aspects of that codified by the SETI community team 23 years later in the `SETI Detection Protocols' (IAA SETI Group 1990) and 37 years later in proposals (IAA SETI Group 2004) for  dealing with sending signals out in the 'Proposed SETI Reply Protocols'.

Almost as though they were following the Detection Protocols, the group investigated the source in detail, with repeated observations, observed it with another telescope and ruled out man-made sources or a hoax. They made further investigations. While they were considering the LGM explanation they did not make any public announcement. They did not follow the recommendation to seek confirmation from another group, but this was a step that perhaps they would have taken if they had ever reached certainty for an LGM origin. They did discuss how they might announce it, and the need then to involve the government and the national academy. Again perhaps because they were only at a preliminary stage, this discussion did not go into details, such as invoking the United Nations, details that might have arisen with an approach to the government.

The concerns raised in the SETI Reply Protocols were considered by Ryle in the pre-Christmas discussion between him and Hewish. Ryle thought that sending a signal out could have disastrous consequences, and once the news got out that there was an LGM out there that it would be impossible to prevent someone sending a signal out. When writing about this in more detail, he said that at least there should be international agreement before any such signalling, in line with the Reply Protocols proposal. 

It has been commented that the Protocols, although desirable, are somewhat theoretical in that experience with `false alarms' has shown that the news leaks out regardless, and any coordination is unlikely to be successful. The Cambridge team were perhaps fortunate that they were able to keep their possible discovery secret. Times were simpler in 1967 - no email, blogs, Twitter, Facebook -  and there were fewer tabloid science journalists then than there are now.

Most present day SETI groups are aware that the nickname `LGM' was used during the pulsar discovery, but the fact of the serious consideration of an ET explanation is less well known. As a result some lessons may  be drawn from it. The Detection Protocols seem to have been valid in this case and so no case for major modification can be made. The secrecy which was sustained for some months was a vital part of the process. If news of the December 21st discussion had got out, the public uproar and misplaced condemnation - `astronomers conceal message from aliens' - would have taken time to sort out and significant damage caused to the group - `astronomers try to scare the public'. One could conclude that, where they are lacking SETI groups need to have plans in place to maintain secrecy, especially in the social networks, for preparing more for a very proactive response if a `false alarm' news starts to leak, and for having detailed government contact and media plans for the event of `Contact'. The response of the media to the news that the group had considered the ET explanation is instructive. Although it was of interest and featured prominently in the reports, it does not seem to have caused much upset. There has been much discussion (see e.g. Billingham et al., 1999) of the possible public response to an ET detection, the pulsar experience seems to be supporting evidence for the judgement that the media and the public would deal with it in a reasonable way. Of course, the detection of an actual message could well lead to more unpredictable outcomes.

The uncertainties in the historical record of the 1967 event are obvious. It would help future historians if SETI groups were to ensure that they have a detailed plan for the systematic recording of events, even those which turn out to be false alarms. The value of such a systematic record in the event of `Contact' would be immense.
 
This was a possible SETI discovery made by a non-SETI group. This could well happen again, when such a group could stumble upon something that looks like ET. This episode should serve as a reminder for the SETI community to continue to spread the fact of the existence of the Protocols as widely as possible amongst all branches of astronomy.

The lessons to be drawn concerning the Response Protocols are more complex. As has been mentioned, no agreement has been reached on how to deal with the prospect of sending a message out, whether that would be a benign, neutral or dangerous act. Papers published since 1960 have come to very different conclusions. But Ryle was certainly right in that if ET is detected, some nation, group or individual will want send a very strong reply back - a signal that would be received by that ET much more strongly than any of our present `normal' radiation. It would also at the minimum alert that ET that we had reached a certain level of science and technology. It would seem to be important to continue to press for an agreed methodology for the response. The scientific communities and the governments of the world need to be prepared. The 1967 episode indicates how difficult it would be to construct a policy in the fervid atmosphere of a `Contact'. In spite of the SETI community's problems in agreeing a Reply Protocol, perhaps Ryle's 1976 approach to the IAU could be tried again.

\section{A NOTE ON SOURCES}

 Although there is considerable contemporary and near-contemporary documentation, the narrative is held together by Hewish's 2008 autobiographical recording. Memories are famously unreliable, but Hewish's account matches in well with the documentation.

The documentation is extensive. There were press articles in March 1968, Bell's 1968 thesis and Hewish's October 1968 article. Ryle's 1968 letter to Woolley and his Trinity talk in 1971 are of particular interest. In 1975 there were the two articles quoted - the Hewish Nobel Lecture and Bell Burnell's ``Petit Four" speech.  Also in 1975 Nicholas Wade wrote an article in `Science', when he appears to have spoken to Bell Burnell, and where he describes (Wade 1975) the period in mid-December as ``when the Cambridge radioastronomy group was seriously considering that the pulses might be signals from another civilization". Woolgar's thesis work in 1975 involved interviews with many of the participants. 

Some of the other material comes from recent recollections by members of the discovery group in personal communications to myself, forty or more years after the event. However, these different memories also chime well, except for those of Bell Burnell, and are also consistent with the contemporaneous matter. It seems reasonably certain that at least some of the group were in mid-December seriously considering the `LGM' hypothesis. 

It is felt that the existing documents and memories give a reliable account of the SETI episode. Is there other material yet to come to light and could further interviews give more details? The Hewish papers in the Churchill archives are very limited. Woolgar's thesis is built on numerous interviews and on Bell's logbook, but he says ``The pursuit of further sources of documentary evidence did little to resolve this ... Notes and logbooks either contained little information or have been lost.".  However, the present contacts with the surviving participants have been by email and perhaps face-to-face interviews could give more, and perhaps a wider search for documents could be successful. Certainly there are a number of other, non-SETI, aspects of this fascinating episode which are still unclear.

\section{A PERSONAL NOTE}

On a personal note, I will recount my own peripheral experience of this event. Early in 1967, I was a final-year undergraduate at Cambridge, interested in astronomy. Each science department puts on a demonstration of their work at that time in the academic year to attract bright students to do a PhD in their subject. So one day I joined a group which cycled out to Lord's Bridge to see the telescopes. We saw the `vineyard-like' array of wires on poles that was the two hectare telescope and a nice little device made by John Pilkington (Pilkington 2011) demonstrating the scintillation effect the telescope would shortly look for. This was a rotating circle of frosted glass in front of a small and a large lamp bulb. The small light changed brightness while the large one did not. It struck me as very clever, but rather boring. I was more interested in optical astronomy, where you could actually see the objects. Which goes to show - what do undergraduates know? I indeed ended up at an optical observatory, the Royal Greenwich Observatory, then deep in the Sussex countryside. One day the next February Richard Bingham, a housemate, asked me how quickly an astronomical source could vary. Thinking about an absolute limit, and remembering that the Sun was a few light-seconds across and there were stars smaller than Sun, I replied `about a second', without thinking about all the practical problems in a star-like object varying at such a rate. Richard was disappointed that I wasn't amazed with this rate, but explained that such a varying object had just been discovered and that the RGO was going to look for the optical counterpart. Following a design proposed by Ryle in a letter (Ryle 1968a) to the Director of the RGO, Bingham built a camera with an image tube behind a semicircular disc rotating at the pulsar frequency. This device and its use were described (Bingham 1968) by him at the April 1968 meeting of the Royal Astronomical Society.

{\bf ACKNOWLEDGEMENTS}

I would like to thank Antony Hewish, Jocelyn Bell Burnell, John Pilkington, Paul Scott, Donald Lynden-Bell and Craig Mackay for their comments, and also to thank Roger Wood and Keith Tritton  for their help with the RGO aspects. Also the help of the staffs of the Churchill Archives at Churchill College,  of the Cambridge University Library, and of the IAU archives are gratefully acknowledged. The author is solely responsible for the contents of this paper.

{\bf REFERENCES}

Bell, S.J. (1969)
The measurement of radio source diameters using a diffraction method.
{\it Ph.D. thesis Univ of Cambridge}, Ph.D. Dissertation 6567 

Bell, S.J., (1977)
Petit Four.
{\it Annals of the New York Academy of Science} 302, 685-689.
(see also www.bigear.org/vol1no1/burnell.htm)

Bell Burnell, S.J. (2008)
{\it BBC programme - Scientists}. \newline
http://www.bbc.co.uk/science/space/universe/scientists/jocelyn\_bell\_burnell\#p009s58j

Bell Burnell, S.J. (2011)
personal communication

Bell Burnell, S.J. and Hewish, A. (2010)
The Discovery of Pulsars.
{\it BBC Four series ``Beautiful Minds" programme (07 Apr 2010)}, \newline
http://www.bbc.co.uk/programmes/b00ry9jq, \newline
( http://www.youtube.com/watch?v=zlDz1xDblWI )

Billingham J., et al. (1996)
Declaration of Principles Concerning Activities Following the Detection of Extraterrestrial Intelligence.
{\it IAA Position Paper, Annexe I},
(also http://www.seti-inst.edu/science/principles.html)

Billingham J., et al. (1999)
Social Implications of the Detection of an Extraterrestrial Civilization,
SETI Press, ISBN 0-9666335-0-4

Bingham, R.G. (1968)
Meeting of the Royal Astronomical Society.
{\it The Observatory} 965, 128 

Burbidge, G. (2007)
An Accidental Career.
{\it Ann. Rev. Astron. Astrophys.} 45, 1-41

Cocconi, G. and Morrison, P. (1959)
Searching for Interstellar Communications.
{\it Nature} 184, 844 

Cornell News (1999)
It's the 25th anniversary of Earth's first (and only) attempt to phone E.T.
{\it Cornell Press release Nov 12., 1999} \newline
http://web.archive.org/web/20080802005337/ - \newline
\hspace*{3em} http://www.news.cornell.edu/releases/Nov99/Arecibo.message.ws.html
 
Drake, F. D. (1961) 
Project OZMA.
{\it Physics Today}. 14(4), 40-46

Drake, F. D. (1965) 
The Radio Search for Intelligent Extraterrestrial Life, 
in {\it Current Aspects of Exobiology}, eds. G. Mamikunian and M. H. Briggs, Pergamon Press, Oxford, p.323.

Graham-Smith, F. (1984)
Martin Ryle. 27 September 1918 - 14 October 1984.
Biogr. Mems. Fell. R. Soc. 32, 496-524

Haqq-Misra, J., Busch, M., Some, S. and Baum, S.
The Benefits and Harms of Transmitting into Space.
http://arxiv.org/abs/1207.5540

Hewish, A. (1967)
Letter to Donald Lynden-Bell at the RGO, 29th November 1967.
{\it Lynden-Bell archive}

Hewish, A. (1968a)
in ``Meeting of the Royal Astronomical Society",
{\it The Observatory}, 965, 128 

Hewish, A. (1968b)
Pulsars.
{\it Scientific American} 219, No.4, pp.25-35 October 1968

Hewish, A. (1992)
Pulsars and High-Energy Physics,
in {\it Nobel Lectures, Physics 1971-1980}, Ed. S. Lundqvist, World Scientific Publishing Co. 
 \newline www.nobelprize.org/nobel\_prizes/physics/laureates/1974/hewish-lecture.html

Hewish, A. (2008)
Autobiographical recording for Cambridge University 26 Mar 2008,
www.dspace.cam.ac.uk/handle/1810/197579

Hewish, A. (2012)
personal communication

Hewish, A., Bell, S.J., Pilkington, J.D.H.,  Scott, P.E., Collins, R.A., (1968)
``Observation of a Rapidly Pulsating Radio Source",
Nature, 217, 709  \newline
(www.nature.com/physics/looking-back/hewish/hewish.html)

Hoyle, F. and Elliot, J. (1962)
A for Andromeda.
Souvenir, London ISBN 978-0285635883

International Academy of Astronautics SETI Permanent Study Group (1990)
Declaration of Principles Concerning Activities Following the Detection of Extraterrestrial Intelligence.
{\it Acta Astronautica}, 21(2), 153-154

International Academy of Astronautics SETI Permanent Study Group (2004)
Proposed SETI Reply Protocols.
http://www.setileague.org/iaaseti/reply.htm

Mackay, C,D. (2008)
in {\it BBC programme - Scientists}. \newline
http://www.bbc.co.uk/science/space/universe/scientists/jocelyn\_bell\_burnell\#p009s58j

Lawrence, W. L. (1959)
Radio Astronomers Listen for Signs of Life in Distant Solar Systems.
{\it New York Times}, November  22, 1959, p.E11

Lemarchand, G.A. and Tarter, D.E. (1994)
Active search strategies and the SETI protocols.
Space Policy 10, 134-142

Longair, M.S. (1996)
in {\it Our Evolving Universe} pp.71-72
Camb. Univ. Press

Longair, M.S. (2011)
The Discovery of Pulsars and the Aftermath.
{\it Proc. Am. Philosophical Soc.} 155, no 2, pp.147-157 

Lovell, A.C.B. (1976)
letter to Blaauw, 21 August 1976
IAU archives

Lovell, A.C.B. (1977)
letter to Muller, 18 January 1977
IAU archives

Mackay, C.D. (2012)
personal communication

McNamara, G. (2008)
in {\it Clocks in the Sky: The Story of Pulsars} pp.39-53
Praxis, ISBN-13 978-0387765600

Michael, D.N. (1960a)
Proposed studies on the implications of peaceful space activities for human affairs. 
{\it NASA Technical Report} CR 55643

Michael, D.N. (1960b)
Footnotes for proposed studies on the implications of peaceful space activities for human affairs.
{\it NASA Technical Report} CR 55640

Mitton, S. (2011)
Fred Hoyle: A Life in Science.
Cambridge University Press, ISBN-13: 978-1854109613

Muller, A.E. (1976)
letter to Blaauw, 17 November 1976
IAU Archives.

Norris, R.P. (2004)
``How to Respond to a SETI Detection''
{\it Bioastronomy 2002: Life Among the Stars}, Proceedings of IAU Symposium 213. Eds. by R. Norris, 
and F. Stootman, Astronomical Society of the Pacific, 2003., p.493

Penny, A. (2012)
Transmitting (and listening) may be good (or bad).
{\it Acta Astronautica}, 78, 69-71

Pilkington, J.D.H. (2011)
personal communication

Royal Swedish Academy of Sciences (2012)
Les Prix Nobel 1974. \newline
http://www.nobelprize.org/nobel\_prizes/physics/laureates/1974/index.html

Ryle, M. (1968a)
Letter to Richard Woolley at the RGO on 9th February 1968,
{\it The Papers of Sir Martin Ryle}
Churchill Archives Centre. Reference GBR/0014/RYLE, J.304

Ryle, M. (1968b)
letter to  Mr Robertson in Spain on 14th March 1968
{\it The Papers of Sir Martin Ryle}
Churchill Archives Centre. Reference GBR/0014/RYLE, J.305

Ryle, M. (1971)
Ten million tons a teaspoonful.
notes for a lecture at Trinity College Cambridge on 2nd Jan 1971,
{\it The Papers of Sir Martin Ryle}
Churchill Archives Centre. Reference GBR/0014/RYLE, G.100

Ryle, M. (1977)
letter to  Mr Brough in New Zealand on 19th January 1977
{\it The Papers of Sir Martin Ryle}
Churchill Archives Centre. Reference GBR/0014/RYLE, J.351

Scheuer, P. (1968)
in ``Meeting of the Royal Astronomical Society".
{\it The Observatory} 965, 128

Scott, P. (2011)
personal communication

Staff at the National Astronomy and Ionosphere Center (1975)
The Arecibo Message of November 1974.
{\it Icarus} 26, 462. 

Sullivan, W. (1976)
Astronomer Fears Hostile Attack; Would Keep Life on Earth a Secret.
{\it New York Times} 4 November 1976, p.46

Tarter, J. (2009)
SETI - Planning for Success: Who Will Speak to Earth? What Will They Say?
in 35th Professor Harry Messel International Science School
http://www.scienceschool.usyd.edu.au/history/2009/media/lectures/10-tarter-chapter.pdf

Wade, N. (1975)
Discovery of Pulsars: A Graduate Student's Story.
{\it Science} 189, 358-364

Woolgar, S.W. (1976)
Writing an Intellectual History of Scientific Development: The Use of Discovery Accounts.
{\it Social Studies of Science} Vol. 6, No. 3/4, pp. 395-422

Woolgar, S.W. (1979)
The Emergence and Growth of Research Areas in Science with Special Reference to Research on Pulsars.
{\it  Ph.D. thesis Univ of Cambridge} Ph.D. Dissertation 10968

\end{document}